\documentclass[12pt]{article}

\usepackage{mathrsfs,amsbsy,amssymb,latexsym,amsfonts,amsmath}
\usepackage[dvips]{graphicx,color}

\global\arraycolsep=3pt
\newcommand\blue[1]{\textcolor{blue}{#1}}

\newcommand\lnf{ln \, f}

\newcommand\bref[1]{(\ref{#1})}

\usepackage{amsbsy}
\usepackage{alltt}
\begin{document}
\begin{flushright}
OIQP-10-04 \\
RIKEN-TH-189
\end{flushright}

\vspace*{0.5cm}

\begin{center}
{\Large \bf 
Backward Causation in Complex Action Model
\footnote{To apper in Proceedings for 13 Bled workshop 
``What Comes Beyond the Standard models", 12-22. July 2010.}
}\\ 
{\large \bf
--- Superdeterminism and Transactional Interpretations ---
} \\
\end{center}
\vspace{10mm}

\begin{center}
{\large
Holger B. Nielsen}\\
{\it
The Niels Bohr Institute,
University of Copenhagen,
\\
Copenhagen $\phi$, DK2100, Denmark
}
\\
and
\\
{\large
Masao Ninomiya} \\
{\it
Okayama Institute for Quantum Physics,\\  
Kyoyama 1, Okayama 700-0015, Japan \\
{\footnotesize and} \\
Theoretical Physics Laboratory,\\ 
The Institute of Physics and Chemical Research (RIKEN),\\ 
Wako, Saitama 351-0198, Japan.}\\
\vspace{7mm}
Keywords: {Backward causation, Superdeterminism, \\
\qquad \qquad \qquad \qquad \qquad \  Transactional interpretation of Quantum Mechanics, \\
 \qquad \qquad \qquad \qquad \ Complex action, Experimental proposal at LHC}
\end{center}

\vspace{1cm}

\begin{abstract}
 It is shown that the transactional interpretation of quantum mechanics being refered 
back to Feynman-Wheeler's time reversal symmetric radiation theory has 
reminiscences to our complex action model. In this complex action model 
the initial conditions are in principle even calculable. Thus it 
philosophically points towards superdeterminism, but really the Bell 
theorem problem is solved in our model of complex action by removing 
the significance of signals running slower than by light velocity.

Our model as earlier published predicts that LHC should have some 
failure before reaching to have produced as many Higgs-particles as 
would have been produced the SSC accelerator. 
In the present article, we point out that a cardgame involving 
whether to restrict LHC-running as we have proposed to test our model will 
under all circumstances be a success. 
 
\end{abstract}

\tableofcontents
\vspace*{0.5cm}

\section{Introduction}

~~~~In the previous articles \cite{search} we proposed that one should use 
the LHC-machine to look for 
backward causation effects. 
Indeed, we proposed a model \cite{search,3,4,5} 
in which the realized history of the 
universe was selected so as to minimize a certain functional 
of the history, 
a functional being the imaginary part of the action $S_I$\,[history], 
which only exists in our model.
In general, it is assumed in science that there is no pre-arrangement \cite{r1} 
of initial conditions so as to make special events occur or 
not occur later.
However J.~Bell proposed in BBC radio broadcast as a solution to the problems 
of Einstein - Podlosky - Rosen's ``super-determinism'' \cite{6} 
and even more it has been developping to the transactional interpretation of 
quantum mechanics \cite{Cramer}, 
which involves Feynman-Wheeler's radiation theory\cite{Wheeler_Feynman_interaction} 
that has backward causation 
in its formalism. 

Also, one of the present authors (H.B.N.) and his group 
earlier proposed models nonlocal in time (and space) \cite{7,8,9}.
Similar backward causation effects have also been proposed 
in connection with the story that e. g. humanity would cause 
a new vacuum to appear, ``vacuum bomb,'' 
by one of the present authors (H.B.N.) and collaborators \cite{10}. 

Our proposal is to test if there should perhaps be such 
pre-arrangements in nature, that is, pre-arrangements that prevent Higgs 
particle producing machines, such as LHC and SSC, from being functional. 
Our model with an imaginary part of the action \cite{search,11}
begins with a series of not completely convincing, 
but still suggestive, assumptions that lead to the 
prediction that large Higgs producing machines should turn out not 
to work in that history of the universe, which is actually 
being realized.

The main points of the present article are the following two points:
\begin{itemize}
\item[A)] To argue that our ``model with imaginary action''[1-4] is a very 
natural type of model if one decides to go for Bell's proposal of 
superdeterminism [6] or the transactional interpretation.
\item[B)] To argue that by making the type of experiment testing our 
model by a card game about the putting restrictions on the running of LHC 
can only seem to be successfull. 
\end{itemize}

These points will be explained below: \\
A)~~~~The point of superdeterminism is to overcome the trouble of the Bell's 
theorem with quantum mechanics and very general assumptions of locality,
by discarding the assumption that the experimentalists in each  of the 
entanglement connected objects  have a ``free will'' to choose what they 
want to measure. Rather the idea is that they could NOT have chosen 
to measure anything different from what they indeed go to measure. I.e. 
it is at least as if the decision of these experimentalists were fixed 
so that it could perhaps even be something that were in principle calculable.
At least it should not be allowed to argue about other possibilities for 
their choice of measurements to perform than the ones they really do choose. 
In that case of course the whole story of the Einstein-Podolsky-Rosen or 
equivalently Bell's theorem has no point. If you only measure whatever you 
measure it is only that element of reality that corresponds to that which is 
relevant and no paradoxes come in.

Since our model leads in the classical approximation to be a model even for 
the initial conditions so that indeed {\em everything} get determined by 
in principle - but not in practice -- from pure calculations only using the 
coupling constants including their imaginary parts as input is an eclatant 
example of superdeterminsim (at laest classsically). We want to argue in 
section 2 below that indeed something like our model is strongly called for
by the problems of measurement theory, as is also stated in the proceedings 
at the Vexjo conference by one of us \cite{14}. 
\\
B)~~~~The plan behind the practical experiment, which we proposed, 
was to produce some random 
numbers--partly by drawing cards and partly by quantum random number 
generation -- and then let these random numbers be translated, according 
to the rules of the game, into some restrictions on the luminosity or 
the energy or both of the LHC. Thus LHC might, for instance, only be 
allowed to run up to a certain beam energy.
I.~Stewart \cite{11} proposed that pauses are determined by random numbers.

The idea is merely to require any restriction at all for LHC  
with probability $p$ that is deemed, by the rules of the game, 
to be very small. 
The probability $p$ for a ``close LHC'' card is \,
$p\sim 10^{-6}$ or so \cite{search}. 

It is clear that even a small probability restriction enforced on LHC, 
its luminosity or beam energy, 
means an artificially imposed -- one would say, ignoring our 
type of model -- 
risk for the LHC project.

It is, 
however, 
the main focus of the present paper to point out 
(as was briefly stated in the previous article \cite{search}) 
that 
even though our proposed project of restricting LHC 
according to random numbers seems to give rise to a loss, 
in fact, whatever happens seems 
-- initially at least -- to be a gain, a success!

That a success in this sense is guaranteed to be the result 
seemingly with almost 100\% certainty (but in reality not quite 100\%) 
is demonstrated in the present article.

The point really is that seeing any restriction coming up from a card game 
of a type that should happen with probability of $10^{-6}$ is 
already evidence for our model, which would thus be discovered by such 
a card draw.

Of course the whole excercise of making the proposed card game experiment 
looking for some backward causation of the type we propose would be futile 
if such an effect were already excluded by earlier experiments. 
If indeed we should look for completely disaterous bad luck for any attempt 
to produce just a single Higgs particle, presumably the Tevatron of 
FNAL in Chicago would already present a counter example. 
Although not a single Higgs particle has been safely recognized to have 
been produced at the Tevatron one expects that according to theoretical 
expectations in say Standard Model several thousands of them would alresdy 
have been produced, although even that is not sufficient for a discovery, 
since only exclusions of mass regions so far were found.
But the LHC accelerator as well as the SSC would, if working, produce much 
more Higgf particles in the long run than the Tevaron. 
So it is certainly a possibility that the effect causing backward particles 
achieved in the LHC and the in 1993 stopped SSC, 
while being insignificant in the Tevatron case.

It has also been proposed that the mere observation of cosmic rays with 
sufficient energy so as to even on fixed target produce Higgs particles 
should represent an argument against the possibility of the effect 
we propose to investigate. However, although we certainly would, 
if such an effect existed, 
predict that the amount of cosmic rays with such 
energies would be reduced by the backward causation effect, one might imagine 
that sources of cosmic rays might be directed to send their radiation 
in the direction of regions with low density of stars and planets so as to 
avoid Higgs production, but we do not have sufficient statistics to have 
any measurement of whether there should --say 300,000 Higgses-- statisically 
be any effect of that type; for that our understanding of what the amount 
of cosmic rays without there being such an effect is too poor.
\section{Relation of Super-determinism to our complex action model}

~~~~We have already remarked in the introduction that 
in our complex action model the imaginary part of the action comes to play 
the role of determining the initial conditions. 
If we indeed denote the complex action 
\begin{equation}
  S\left[\mathrm{path}\right]=
    S_R \left[\mathrm{path}\right]+iS_I \left[\mathrm{path}\right],
\end{equation}
where $S_R$ and $S_I$ are then the real and imaginary parts respectively 
we have approximately that among all the solutions to the classical equations 
of motion in ignoring the imaginary part approximation
\begin{equation}
   \delta S_R =0, 
\end{equation}
the solution with the minimal (i.\,e.\ most negative) $S_I[\mathrm{sol}]$ is 
the one which we live through (i.\,e.\ the one realized). 
That is to say the formula in our complex action model for the history of 
the universe to be selected as the one realized becomes 
\begin{equation}
   S_I \, [sol\,]\, \mathrm{will\ be\ minimal}.
\end{equation}
The fact that we have such a formula -- wherein the mathematical expression for 
$S_I$ in terms of the field(development)s is very similar and analogous to 
the usual Standard Model action expression, except that the coefficients 
deviate -- for the realized history $sol$ means that even the initial 
conditions (contained in $sol$) are \emph{calculable} in principle, 
although not in practice.
With such a model as ours in which one thus can calculate ``everything'' 
in principle one can especially imagine calculating the choice of the 
experimentalists in an 
EPS(=\,Einstein-Podolsky-Rosen) type of 
experiment would perform on the particles that were separated in this experiment. 
Let us remind the reader that in the EPS or Bell-theorem type of experiment 
a couple of quanta(=\,particles) are produced in a correlated (entangled) state 
and successively these particles separate to run in different directions. 
 Further away whereto these separated particles run two different 
experimentalists teams with their detections determine measure properties 
of the particles. 
It is important that these experimentalists, if they have ``free will'', can 
make their decisions about what properties to measure, 
the spin component along what direction say, or momentum or position, 
\emph{after} the particles are already widely separated. 
 Of course in our model with complex action, like in even just usual 
deterministic models, the ``free will'' will only be something in our fantasies, 
since we could in our model in principle even have calculated what their choices 
of quantities to be measured would be.

 The problem with quantum mechanics associated with 
Bell's Theorem is that under mild assumptions, mainly that no signal 
can go arbitrarily fast from the one measurement place to the other one 
so as to communicate the experimentalists choice, the quantum mechanical 
statistical predictions are not consistent. 
 If one takes the point of view that we only need to consider the truly realized 
situation, i.\,e.\ choice of experiments measuring on the particles,  
but can ignore totally the ``only fantasy'' possibilities associated with 
the make up of a ``free will'' feeling for the experimentalists, then 
Bell's theorem falls away and quantum mechanics has no Bell's theorem problem. 

 Thus logically we may say that our model even strengthens the application 
of superdeterminism to escape the Bell's theorem problem. We say that our model 
strengthens the superdeterminism because it makes the initial state 
and thus the deterministically determined experimentalist decisions 
even calculable, so that a requirement that we should be allowed to vary 
at least the initial conditions in deriving Bell's theorem requirements would 
no longer hold in our model. 

 Now, however, we shall argue in the next section that in spite of 
our model in this way strengthening the ground for superdeterminism, 
it is in fact another feature of our model that removes 
the Bell's theorem troubles in it.
 In fact the point is rather that once we have effectively backward causation 
-- so that e.\,g.\ the potential switch on of SSC to produce 
many Higgs bosons can \underline{backwardly} cause 
the Congress of the United States to stop the funding -- 
then the rule that a signal cannot move with arbitrarily high speed 
no longer makes sense.
 The signal could instead move slowly along in the future and then go 
backward in time using the feature of backward causation in our model. 

\section{Analysis of the way our model solves the EPS-Bell's Theorem problem}

~~~~Really as Bell himself were aware
-- and why he thus did not like superdeterminism as the way 
out for quantum mechanics -- 
there can be a lot of small details going on in an in practice 
very hard to control way and these details can influence the experimentalist's 
decisions. Such details are e.\,g.\ the reasoning in their brains and 
really may well represent their ``free will''. We believe that indeed a 
superdeterminism solution solving the Bell's theorem problem for quantum mechanics 
by postulating that even these ``free will'' -- simulating details leading 
forward to the experimentalist's decisions can be somehow integrated up 
and calculated through by the particle choosing its property when measured 
is a somewhat unhealthy philosophy. 
 How should indeed a particle say B at site B where a measurement is 
made ``know'' and ``understand'' the contemplations in 
detail of the experimentalists at the other site A\,?\ 
 It sounds healthier to make an assumption that such detailed calculations 
as the ones in the experimentalists of team A cannot get calculated through 
at the observation site B. Supplementing quantum mechanics with locality 
etc by such a reasonable extra postulate the loophole in the Bell's theorem trouble 
for quantum mechanics would be closed. With such an extra reasonable assumption 
quantum mechanics would be truly in trouble. 
 Our strengthening by making the initial state ``calculable'' in principle will 
a prior not help much against the reasonable assumption suggested. 
 So as far as the true superdeterminism solution to the Bell's theorem problem 
for quantum mechanics our model does not help much. 

 There is, however, \underline{another way} in which our model may help more realistically quantum mechanics against the Bell's theorem problem. 
 Since the formula $S_I \mathrm{[history]_{minimal}}$ has in the integral form 
\begin{eqnarray}
   S_I \left[\mathrm{history}\right]
   =\int_{\mathrm{all~times}}L_I \left(\mathrm{history}(t)\right)dt
\end{eqnarray}
contributions from all times, from the beginning of times to the end, it also 
includes contributions from what is future for us to say. 
 In order that our model shall have a chance of being viable we must of course 
hope or speculate that e.\,g.\ because of the special conditions in the 
inflation time, the contributions to $S_I [\mathrm{history}]$ from 
$L_I \left(\mathrm{history}(t)\right)$ for time $t$ in the inflation era were 
by far the most important, so that what happened in the inflation-era 
dominated the selection of what history were the one to be realized 
(to be the one we live through now). 
 Only with such an assumption of the inflation-era contribution to the 
$S_I =\int L_I dt$ \ integral dominating the selection of the initial 
conditions (the solution) will agree with our normal experience with second law of 
thermodynamics, meaning that only the start were strongly organized 
in the sense of having low entropy and essentially nothing being prearranged 
by having fine tuned initial conditions destined to make future thing happen. 
 However, in our complex action model there should be at least some seeming 
attempts to such prearrangements, meaning that there is in our model in principle 
happening events not expected statistically in the usual theory, so that one 
might denote them as ``miracles'' or ``antimiracles'' if it is something bad. 
 Usually, however, we expect that the contributions from the era such as the 
inflation era around the ``Big Bang'' time (if there were a Big Bang) 
would dominate. 

 When, however, we consider a quantum experiment with a measurement, 
the result of which seemingly independently of the initial conditions can with 
finite none-zero probabilities obtain different measured values, it becomes 
suggestive that the future contribution 
$\int_{t_\mathrm{exp}}^{\infty }L_I dt$ 
could become important where text denotes the time the experiment is performed. 
 We therefore in our model suppose that the outcome of a quantum experiment 
is not a priori just pure fortuitousness or accident, but actually depends on 
the (future) contribution to the imaginary part of the action 
$\int_{t_\mathrm{exp}}^{\infty }L_I dt$. 
 That is to say, we expect that the outcome of the measurement is that result 
which minimizes the contributions 
$\int_{t_\mathrm{exp}}^{\infty }L_I dt$\, 
to
$S_I$ depending on this outcome. 

 If we have, as we now assume about our model, a theory in which 
the outcome of a quantum measurement is selected by minimizing an integral 
$\int_{t_\mathrm{exp}}^{\infty }L_I dt$  
extending into the far away future for the whole world, then the worry about 
signals going between the sites A and B for the measurements in the EPS-type 
experiment faster than light or arbitrarily fast looses its interest. 
 The point namely is that the to avoid Bell's theorem troubles needed 
faster than light signals can be replaced by signals reaching the future 
of the particles measured upon, because it is the future of the particle 
(roughly speaking) that determines the result of the quantum measurement. 
 Really it is not so much the future of the particle itself as of the results 
of the measurement as propagated by publications etc into the future 
that matters for the 
$\int_{t_\mathrm{exp}}^{\infty }L_I dt$  
integral to be minimized for telling the result of the measurement.
  You might describe this picture of the measurement results being determined 
by minimizing the future part of the imaginary part of the action $S_I$, 
called $\int_{t_\mathrm{exp}}^{\infty }L_I dt$, 
crudely as the information first going forward in time where it all the way 
contributes to $\int_{t_\mathrm{exp}}^{\infty }L_I dt$ \, and 
cause what we call backward causation an influence backward in time 
determining the measurement results. 
 
 To make clearer the way our model treats the Bell's theorem situation, 
we could hypothetically imagine that the two sites of measurement A and B 
were kept in all the future isolated. Then the measurement results, 
in our model determined (in a complicated way) from the future integral 
contribution $\int_{t_\mathrm{exp}}^{\infty }L_I dt$, 
could not get correlated.
In other words, the anti Bell's theorem strange correlation 
(or any correlation) 
between the measurement results provided the two sites A and B have a common 
future, which can contribute to $\int_{t_\mathrm{exp}}^{\infty }L_I dt$ \,
and thereby make the minimization of this integral provide the correlation. 

 This ``explanation'' of the violation of Bell's theorem by the future 
contributions to $S_I$, i.\,e.\, 
$\int_{t_\mathrm{exp}}^{\infty }L_I dt$, 
deciding the measurement results in our model is nicer than the genuine 
superdeterminism, because it does not require the complicated contemplations 
of the experimentalist teams to be ``known'' and ``understood'' by any particles. 

 Really the usual Copenhagen interpretation (or Born) rule is approximately 
 reproduced in our model by making an approximation 
\begin{equation}
    \mid \mathrm{B}(t)\left\rangle \right\langle \mathrm{B}(t)\mid \sim 1, 
\end{equation}
where the ket $\mid \mathrm{B}(t)>$ and its bra $<\mathrm{B}(t)\mid$ is 
given by a functional integral over the exponentiated action from the after 
$t_{\mathrm{exp}}$ era only 
\begin{eqnarray}
   S_{\mathrm{after}\,t_\mathrm{exp}} = \int_{t_\mathrm{exp}}^{\infty} Ldt.
\end{eqnarray}
I.\,e.\ we defined previously in a basis consisting of basis vectors 
$|\vec{q}>$, 
\begin{eqnarray}
  \left\langle \vec{q}\mid \mathrm{B}(t)\right\rangle \hat{=} 
  \int_{\mathrm{with~conditions~path}(t)=\vec{q}} 
  e^{\frac{i}{\hbar}S_{\mathrm{after}(t)}(\mathrm{path})} 
  &&D\mathrm{path}. \\
  &&{\tiny (\mathrm{paths~from~path}~t_{\mathrm{exp}}~\mathrm{to}~\infty )} \nonumber
\end{eqnarray}

 We thus see that although everything even what really happens and what gets 
measured is in our model (super) determined in the sense that it is even 
in principle calculable, the important way in which our model can be claimed 
to solve the problem with Bell's theorem is rather that it by having dependence 
on the future via the integral 
$\int_{t_\mathrm{exp}}^{\infty }L_I dt$ 
gets information/\,a signal backward in time to determine the outcome of the 
measurement.
After such a backward in time signalling is allowed, the locality principle 
formally implemented by the (complex) action being of the form 
\begin{eqnarray}
   S\left[\mathrm{path}\right]=
   \int L\left(\mathrm{path}(x), \partial \mathrm{path}(x)\right)\sqrt{g}dx,    
\end{eqnarray}
where the Lagrangian density 
$L\left(\mathrm{path}(x), \partial \mathrm{path}(x)\right)$ 
only depends on the field development called path in the infinitesimal 
neighborhood of the spacetime point $x$, i.\,e.\, on $\mathrm{path}(x)$ 
and its first derivatives $\partial \mathrm{path}(x)$, 
can still be compatible with an effective arbitrarily fast information transfer. 
It is by means of backward causation via the 
$\int_{t_\mathrm{exp}}^{\infty }L_I dt$ dependence our model 
-- in a somewhat reasonable thinkable way -- circumvents one of the assumptions 
behind the Bell's theorem and thus its trouble for quantum mechanics. 
\section{The Trasnactional Interpretation}

~~~~There is another proposal for quantum mechanics interpretation, 
which is even more similar to ours than the just discussed superdeterminism, 
and that is the transactional interpretation 
In fact this transactional interpretation has formally the interesting 
common feature with our model: Formal influence from the future. 

In the transactional interpretation this formal backward causation or 
influence from the future is at least clearly alluded to by the fact 
that the transactional interpretation on the Feynman-Wheeler electrodynamics. 
In this Feynman-Wheeler theory of electrodynamics the usual boudary 
conditions used to derive electromagnetic radiation to be described 
by retarded waves is replaced by a time reversal invariant boundary condition.
This Feynman-Wheeler postulate is that an electrically charged object 
sends out both a retarded \underline{and} an advanced contribution to the 
electromagnetic fields so that the total emission is time reversal invariant. 
This means that formally fields propagate \underline{both} backward 
and forward in time. Thus formally the Feynman-Wheeler theory has influence 
from the future built into it. It is nontrivial in their theory to argue 
that in practice we obtain seemingly only the retarded waves, and the 
argumentation does not work in all cosmological pictures. 
It is namely based on a discussion in which the absorber of the light is 
strongly needed. 

When therefore the transactional interpretation is based on the wave 
function for the photon ($\sim$essentially the electromagnetic field) is 
influenced by charged matter in just the way proposed by Feynman-Wheeler 
it looks a priori as if the transactional interpretation is also easily 
going to contain influence from the future. 
However, it is claimed by the proponents of the transactional interpretation 
by Cramer 
\cite{Cramer}
that one can distinguish a strong and weak principle 
of causality. The weak principle of causality which only claims that a 
cause shall come before the effect when it applies to macroscopic observations 
and observer--to-- observer communication. 
But Cramer claims: there is no present experimental evidence in support of 
any causal principle stronger than the ``weak principle". 
To this interpretation even opens up for backward causation on the microlevel, 
since strong causality is not hold up. 
Another point pointing towards our model of complex action is the occurrence 
of \underline{two} wave functions: OW (``offer wave") and CW (``confirmation wave"). 
This has similarities to the $\left\langle q|A(t) \right\rangle$ and  
$\left\langle B(t)|q \right\rangle$ wave functions defined in our functional 
integral based on ``complex action model" by the following ``half time" 
functional integrals 
\begin{eqnarray}
   \left\langle q|A(t) \right\rangle =
   \int_{\mathrm{{with~boundary \atop conditions} \atop q'=path(t).}} 
   e^{\frac{i}{\hbar} \int_{-\infty \mathrm{(beginning)}}^{t} \mathrm{L(path,\partial path)}dt} \mathrm{Dpath_{(half)}},
\end{eqnarray}
and
\begin{eqnarray}
   \left\langle B(t)|q' \right\rangle =
   \int e^{\frac{i}{\hbar} \int_{t}^{\infty } 
   \mathrm{L(path,\frac{dpath}{dt}}dt} \mathrm{Dpath_{(half)}}.
\end{eqnarray}

\subsection{4.1 More review of transactional interpretation}
~~~~As far as we understand the point of the transactional model is that 
echoes of advanced waves responding retarded and advanced considered in a 
pedagogical time finally leads to a total field which obey:\\
\begin{itemize}
  \item[a)] the usual type of boundory condictions of no wave before emission and no wave after absorbtion.
  \item[b)] some quantization condictions, e.g. supposedly that the energy is given by a Planck quantization rule.
\end{itemize}

We must think about it that the field being nonzero region gets concentrated 
along a narrow track in space(time) connecting the emmitter to the absorber. 
If this is a correct interpretation of the transactional model then we see 
that the direction of motion of the emitted photon is from the start geared 
to reach its absorption place, the absorber. 
But that means that it is indeed strongly influenced by the future. 
This is of course what is expected in a model based on the backward causality 
containing Feynman-Wheeler theory. It means, at least, that in principle now 
the influence from future has sneaked into the transactional interpretation scheme, 
then it may turn out to not be there macroscopically at the end though. 
\subsection{4.2 Is our complex action model equivalent to the transactional interpretation?}
~~~~Although both our model of complex action and the transactional interpretation 
model are both characterized by influence from the future, 
they are not exactly the same, since we have different details for the influence 
from the future. In fact, there is in our model in principle a series of parameters 
in the form of the imaginary part of the action 
$S_I =\int L_I dt$ to be chosen, before we have a definite model, while in 
the transactional interpretation model one uses the Feynman-Wheeler time reversal 
symmetric emission-rule ($\sim$ boundary condition) to tell 
\underline{how} future influences past.

But in a general way we may bring the correspondence between the two models 
to be very close each other indeed. Presumably the best way to make the correspondence 
be there is to use the second quantized theory in the field theory language in the 
functional integral taken as fundamental in our model.

That is to say, we take our abstract ``path" to mean a thinkable develop 
of all the fields (supposed for simplicity only bosonic fields 
$\psi (X^{\mu })$). 
This means that the phase space -- in this thinking on our model -- 
is a space of infinitely many dimensions the coordinates of which are 
partly the fields $\psi_i (\vec{X})$ and partly their conjugate fields. 

Now it is the crucial feature in our model with complex action and 
use action integral over all time (including both all past and 
all future) that the initial conditions or rather a solution to 
the equations of motion gets field (and is in principle calculable). 
This classical solution singled out by means of the imaginary action 
$S_I$ being minimal is a classical solution describing a path through 
all times. 
It thus even in principle makes it possible to calculate the outcome 
of quantum experiments (in our complex action model). 
Thinking upon this model with the fields as the variables describing 
the path we get thus our model to -- up to a few small splittings of 
the track -- deliver as in prinsiple (but not in practice) a classical 
solution to the field development. 
But now such a classical field development is what from the only quantized 
point of view is a specific development of the wave function. 
This wave function now can be considered as it is in the transactional 
interpretation just an ordinary (meaning well-defined classical) field! 
In this way you can say that our complex action model taken as a theory 
for the fields deliver just the picture of the transactional interpretation. 

\subsection{4.3 How comes single quantization about?}
~~~~With such a making the wave function or say better the fields become 
classical solutions one might become worried about how we can get say the 
quantization of the energy of a photon by $\hbar \omega $, where $\omega$ 
is the frequency.

In our model it cannot really come about unless we allow that there typically 
will be more than just classical solutions selected, but rather a discrete 
series of rather close to each other solutions. 
In the case of a photon being transmitted from some emitter to an absorber 
over a long (space and) distance these close to each other but different 
classical solutions --still relevant/contributing to our functional integral-- 
would be solutions within a range of close by numbers of turns in the oscillations 
of the field from emitter to absorber. But now each extra turn in the 
field oscillation will give an extra phase factor in our ``fundamental" 
functional integral. These different ``neighboring" routes will only add up 
constructively provided the total phase difference between the contributions 
from the different classical (field) solutions happen to be 
(at least approximately) zero. 
Such a condition for constructive interference between contributions in the 
Wentzel-Dirac-Feynman integral from various only a but from each other deviating 
classical solutions could lead to the quantization rule in the single quantized 
language.

It seems that in the transactional interpretation the quantization of energy 
and momentum is imposed as an extra condition without any explanation behind it. 

That would correspond to our model if one would make quantization without 
having our functional integral on a level more fundamental, then namely 
the phase-factor from the behind functional integration 
would  have no place in the picture. One would have to put it on extra 
as a kind of Bohr-quantization condition. 

\subsection{4.4 What to conclude from the tight connection of our model with transactional interpretation?}
~~~~One can look at the close coincidence of our complex action model 
with transactional interpretation in two opposite directions: 
Believing transactional interpretation and show that it is a model 
``of our type" thus supporting such models. Or one could oppositely 
believe in our model and say that derive not exactly the conventional model of transactional interpretation, but a transactional interpretation type model. 
The latter does not necessarily have the Feynman-Wheeler's specific way of sending 
equal strength wave retardedly and advanced, but which in the important 
``philosophical" aspects would be just the same: The wave functions 
(in the single particle picture) could be considered ordinary ($\sim $classical) 
fields, there would be influence from the future so that a particle 
would be guided in the right direction from the start in say 
Renninger's negative result experiment.


\section{What we need}

~~~~Even though it is not so much the superdeterminism in our model in the 
sense of everything being calculable in principle that makes it 
compatible with the Bell's theorem 
and quantum mechanics as its lack of information only going forward in time 
as usual that causes the compatibility of our model with quantum mechanics 
and locality our model is nevertheless supported by the troubles of quantum 
mechanics. 

 We could generally state that clearly any theory with backward causation 
like our model would potentially be able to circumvent the Bell's theorem troubles 
by via the future forth and back allowing an effect/a signal to go effectively   
faster than light. 
Such potential theories with backward causation it would be able to solve 
the Bell's theorem trouble. This type of ``theory'' could be claimed 
to be supported by the quantum mechanics Bell's theorem trouble. 
 This fact makes it especially important to look for any backward causation 
effects whenever there should be a chance for it. Since we have so far only 
rather weak evidence for if any even very seldom prearranged events it seems that 
usual daily life physics should show extremely little backward causation in any 
viable physics theory. However for much higher energy per particle than 
in daily life physics we may have yet looked less carefully for prearranged 
events ($\sim$ miracles).
It is therefore to be suggested that, e.\,g.\ to look for a possible way 
out of the Bell's theorem problem, one should at each new accelerator look for 
prearranged events. 

 If the prearranging governing (e.g.\ via the initial conditions) of the world 
were made to arrange for or arrange for avoiding some phenomenon happening 
due to high energy accelerators of some sort, the easiest 
(least miraculous) way to arrange for or avoid such a phenomenon might be 
to favour or disfavour the very building of the accelerator. 

 As the example which is favoured by speculations in our model of complex action 
it could be that there is a special type of particle which if produced will 
contributes especially much to e.g.\ disfavour the accelerator producing it. 
If so then the type of accelerator producing this type of particle 
-- especially if in large numbers -- 
should be prearranged not to come to work for long time in the mode producing 
the many such particles. In our model, we suggest that the type of particle 
causing the disfavour and giving thus especially bad luck for the running of 
the accelerator is the Higgs boson, because we think that the term 
$\ldots+m_h^2\mid_I \cdot \mid\phi_{H} \mid^{2}+\ldots $  
in the imaginary part of Lagrangian density 
\begin{eqnarray}
   L_I (x)=\ldots+m_h^2\mid_I \cdot \mid\phi_{H} \mid^{2}+\ldots, 
\end{eqnarray}
is dominant from a dimensional argument. 
 The imaginary part of the Lagrangian density $L_I (x)$ is of course the space 
time density for the imaginary part of the action 
\begin{eqnarray}
   S_I \left[\mathrm{path}\right]=
   \int_{\mathrm{over~all~space~time~incl. \atop past~\underline{and}~future}}
   L_I \left(x,\mathrm{path}(x)\right)\sqrt{g}\,d^4 x.
\end{eqnarray}

 Our ``dimensional argument'' is that if the natural units were the Planck units 
the natural value for quantity \,$m_h^2\mid_I$ having dimension of mass square 
would be the Planck mass $m_{\mathrm{PL}}\sim 10^{19}GeV$ squared, i.e.\, 
$m_h^2 \mid_I \sim \left(10^{19}GeV\right)^2 \sim 10^{38}GeV=10^{32}TeV^{2}$ 
which is tremendously large from the point of view of LHC-physics. 

 If an accelerator indeed has the potentiality of producing many of such 
``hated'' or bad luck giving new particles we might observe it by investigating 
statistically if the accelerators meet bad or good luck technically and politically. 
 Here immediately the reader should think of the biggest potential (putative?) 
accelerator the SSC having been stopped in 1993 by the Congress.

 As we have already suggested in earlier papers it might be difficult to get 
a clean statistical investigation of the potential bad luck unless one makes 
a very clean experiment by betting a card game preferably combined with quantum 
random numbers decide whether a certain accelerator 
-- of course we propose it to be LHC -- be brought to run and at what luminousity 
and energy.   

\section{Card game for LHC restrictions can only be a success!}

~~~~There are two possibilities.
\begin{enumerate}
\item[1)] You draw a card combination of the most common type leading to no restrictions.
      Then LHC can run without any restriction and you can be totally happy 
      because you found, with close to zero expense, an argument 
      against our theory.
      You almost kill our theory, 
      or at least drastically diminish the chance that it is right. 
      This is a very good scenario!
\item[2)] You draw a restriction card combination. 
      Now, it is a significant loss that LHC cannot run in full, but
      now you have proved our, or a similar, backward causation theory.
      This would be so interesting, if one really had backward causation, 
      that it might be counted as a discovery greater than supersymmetric partners 
      or the finding of the Higgs.
      It would be a fantastic discovery made with LHC! 
      If the restriction drawn is not a total closing, you would likely soon also 
      find the Higgs and perhaps the supersymmetric partners 
      even if 
      statistics might initially be a bit worse than hoped for. 
\end{enumerate}

 It would be a wonderful victory for CERN and LHC to find 
\underline{backward causation} 
together with having to obey the most likely very mild restrictions.
 We should remember that the rule of our card game should be 
to make the milder restriction have a much higher chance 
of being drawn than the very strong restriction of, for example, totally closing LHC.

Quite correctly, there is, though a little chance of, 
a true loss even though it will not be initially noticed. 
It is possible, although not likely, 
that a random number game leads to a restriction even if our model, 
and any model with backward causation, is wrong.
In this case, we have a bad bargain:
not only would we loose the full applicability of LHC, 
but we would also have gotten, by a statistical fluctuation, the wrong impression 
that a backward causation containing model were indeed true without 
this actually being the case. 

We should certainly arrange the restriction probability $p$ to be low 
enough to make this bad case have a very low probability.

One would, from this way of arguing, initially suspect 
that it would be most profitable not to perform our random number 
LHC restricting experiment 
because
if our theory were right LHC would, in any case, be closed or restricted somehow 
by prearranged bad luck,
as happened to SSC, 
for which Congress in the U.S.A. terminated economic support.
Now, 
however, 
we want to argue that it would be more agreeable to have LHC be stopped 
or restricted by a random number game rather than by some bad luck 
such as political withdrawal of support. 
The main reason for the artificially caused random number withdrawal being 
preferred is that we would, in this case, get more solid support 
for our, or a similar, model being true than by the same restriction 
coming about through a bad luck accident.

To see that would be more convincingly shown 
the truth of our theory of imaginary action determined 
by history  
if we have a card or random number closure rather than a ``normal'' failure, 
we could contemplate
how much more convincing our theory would have been today 
if the SSC-machine had been closed after a random number experiment 
rather than mainly for economical reasons or perhaps because of 
the collapse of the Soviet Union, 
which made the competition 
with 60 million dollar accelerators not worthwhile. 

Now it is sometimes explained that SSC \cite{12} 
had bad luck because of various stupidities or accidents, 
but had it been a card game such ideas would not matter. 
Everything is an accident, 
but we would know the probabilities very reliably. 
So if the card game were set up so that the closing probability 
were sufficiently small, we would have been sure that the closing of 
SSC were due to a (anti)miracle. 

In the following, we shall present a little calculational example to 
illustrate formally that a more reliable knowledge of the truth of 
our theory is obtained with a random number experiment. 
This comes under the discussion of point 2) among the reasons for conducting 
our proposed experiment later in the present article.

\section{Reasons for conducting our proposed experiment}

~~~~What could be a reason to conduct the card game experiment? 
\begin{enumerate}
\item[1)] To obtain greater conviction about the truth of our theory  \\
   -- if it is true of course. -- 
\item[2)] To perhaps avoid bad backward causation effects. 
\end{enumerate} 
These are the two benefits you could have.

In formula it would mean that we should estimate averages 
for the two measures of these two benefits. 

\subsection{More conviction of truth of our model} 

~~~~For reason 1) -- the conviction about our theory 
that it is indeed right -- we need some measure. 
Both the result of the card game and the failure of the LHC 
for other reasons are statistical events, 
but, while we have very trustable ideas about what probability 
$p$ 
to assign to a given class of card combinations, 
our assignment of a trustable value for the failure probability 
$f$ 
for other reasons is very difficult and has a huge uncertainly. 
Therefore, if LHC fails for a reason other than a random number game, 
we would have not even truly learned that our theory was right 
even though we would say 
``it is remarkable that the present authors wrote about the failure 
while LHC still looked to be able to work.'' 

\subsection{Miraculocity and estimating evidence for our model}

~~~~In order to understand why the difference between getting 
our model supported by a ``natural '' failure of LHC and a failure 
caused merely by having a card game drawing a 
``restrict LHC'' card gives rise to an important difference in trustability 
in our model. 
 We shall give a slight formal illustration using the statistical model 
which is not very exact but is appropriate for illustrating our point. 

 If, in our model, a seemingly other reason for failure of LHC 
occurs merely through the coincidence of a series of small bad luck events  
-- that by themselves can easily happen -- 
then the number and unlikeliness of elements in this series of 
bad luck events must be proportional to 
$  - \lnf = | \lnf| $
where
$f$
is the probability of failure. 
We could call this quantity
$ - \lnf $ 
the ``miraculocity'' for failure in a seemingly natural way. 
This concept of 
``miraculocity''
becomes a measure for how many 
``submiracles''
must occur. 
As examples of submiracles, there are  
``the watch man having drunk a bit too much'', 
``the connection between super conducting cables 
  having too high resistance'', 
``The accident being in the difficult part of the tunnel, 
  just under Jura mountain'' etc.

Now if we set up a card or quantum mechanically based 
random number generator leading to 
``restrict LHC''
with probability  $p$ , 
it needs to generate 
-- by the selection of the realized history in our model -- 
a number of adjusted accidents (or submiracles) 
in a number proportional to 
$ -ln\, p =|ln\, p| $ .
Essentially, in the case of the truth of our theory, whether the failure 
of the LHC will arise via the card or the quantum random number game 
or via a natural reason will depend on which of the two alternative 
miraculocities 
$ -ln\, p $
or 
$ -ln\, f $
is the smallest. 
There will, of course,  be a preference with ``miraculocity'': 
the least miraculous of the two alternative possibilities for failure will 
most likely be the one that occurs. 
This would require fewest submiracles. 

We can define $f$ so that indeed 
$-ln\, f$ 
gives a measure of the ``miraculocity'', but it is very difficult 
even for people building the LHC, to convincingly figure out 
what to accept or predict about this miraculocity 
$ -ln\, f =|ln\, f| $ .
At best, one can predict it with an appreciable uncertainty. 
That is to say, we obtain, at least from some simulation 
-- say by Monte Carlo methods or  
just theoretically -- a probability distribution for 
``miraculocity'' $|ln\, f|$. 
 To illustrate our point of estimating the degree of 
conviction, which we shall obtain in the case of a 
``natural'' and\,/\,or ``normal'' 
failure, we can assume that the probability calculation -- by (computer) 
simulation of the political and technical procedures around 
CERN and LHC -- led to a Gaussian distribution for the miraculocity 
$-ln\, f$.
That is to say, we assume the probability distribution
  \begin{eqnarray}
     && P\left(| ln\, f | \right) d\,|ln\, f|  =  \nonumber \\ 
     && \approx \frac{1}{\sigma \sqrt{2\pi  }}\> \mathrm{exp} 
        \Big( \frac{1}{2\sigma^2}\left(|ln \, f|-\ |ln \, f_0 |\right)^2 
        \Big) \ d\,|ln\, f|.\nonumber \\ 
      \label{Gauss}
  \end{eqnarray}
Here, $\sigma$ is the spread of the distribution for the logarithm of 
$f$ , i.e., 
the ``miraculocity.'' 

Now let us consider the degree of remarkableness for the failure 
depending on whether it is due to the card 
or the quantum random number game 
or a ``normal'' failure, i.e., other reasons 
such as meteors and bad electrical connection between the superconductors. 

In the case of a card or quantum random number game, the number of 
sub-miracles in the card or quantum packing is proportional to 
$-ln\,p$, 
where $p$ is the arranged probability by the game rules. 

However, if there is instead a ``normal'' failure due to the stupidity of some 
members of cabinet or the like, then we would tend, of course, 
to believe that the true miraculocity 
$-ln\, f = |ln\,f|$
for that failure is indeed in the low end of the estimated Gaussian 
distribution.
In other words, we would expect that, after all, the ``true'' probability 
for failure $f$ is rather high, i.e., 
$f>f_0$ 
or presumably even 
$f \gg f_0 $ 

Let us indeed evaluate the expected probability for a seemingly 
``normal'' (i.e., not caused by card etc games) failure. 
This expected normal probability for failure is 
\begin{eqnarray}
\left\langle f \right\rangle
  =\int ^{\infty }_{-\infty }\frac{1}{\sigma \sqrt{2\pi }}\cdot f 
   \cdot \mathrm{exp} \left(-\frac{1}{2\sigma ^2}(ln\, f-ln\, f_0)^2 \right)
d\,|ln\, f|
\end{eqnarray}
(we imagine that the miscalculation by including the 
$f>1$ region is negligible, but one could of course do better if needed). 

  We immediately write 
$f=e^{-|ln\,f|}$. 
We had hoped to expect ``normal'' failure 
with the probability given by 
\begin{eqnarray}
\left\langle f \right\rangle 
    &&=\int ^{\infty }_{-\infty }\frac{1}{\sigma \sqrt{2\pi }}
       \cdot \mathrm{exp} \left(-\frac{1}{2\sigma ^2}(|ln\, f|-|ln\, f_0|)^2 
       -|ln\, f| \right)d\,|ln\, f| \nonumber\\
    &&=\int ^{\infty }_{-\infty }\frac{1}{\sigma \sqrt{2\pi }} \cdot \mathrm{exp} 
       \biggl(-\frac{1}{2\sigma ^2}\left[ \left(|ln\, f|-|ln\, f_0|
       +\sigma ^2\right)^2  \right. \nonumber \\
    && \qquad \qquad \qquad \qquad \qquad \qquad \qquad \quad      
         \Bigl. -\sigma^4 +2\sigma^2 |ln\,f_o| \Bigr] \biggr) d\,|ln\, f| \nonumber\\
    &&=\mathrm{exp}\left( \frac{\sigma^2}{2}-|ln\, f_o|\right) =f_o\,e^{\sigma^2 /2}. 
\end{eqnarray}

 Hence the remarkability or apparent miraculousness of the 
outcome that LHC should fail seemingly by a ``normal'' accident 
-- such as political closure -- 
is not the ``miraculocity'' corresponding to the most likely value for 
$f$, i.e., $-ln\,f_o =|ln\, f_o |$, but rather to 
$\mathrm{``remarkableness''}=-ln\left\langle f \right\rangle =|ln\left\langle f \right\rangle\!|=|ln\,f_o|-\frac{\sigma^2}{2}$. 

It is this correction by the term 
$-\frac{\sigma^2}{2}$ 
that causes less conviction for our model being true 
if the failure of LHC shows up as a ``normal'' failure, 
than if we get a failure caused by a card or quantum random number game. 
One should keep in mind that whether in our model one or the other 
reasons for failure occurs depends largely on the relative sizes of 
$-ln\, f$ and $-ln\, p$ . 

 In this way, it would be more convincing that 
our theory were true if the failure were found by a card game or the like than by a 
``normal'' failure of LHC.
 It would thus be profitable scientifically if we could provoke a 
card game failure instead of a ``normal'' one; we would have 
the possibility of arranging that if our model were right. 
 In the case of our model being wrong, of course, the card game project 
would only add to the totally failure probability of LHC, 
making a card game a risk and a bad thing.

 Should our theory be right, the failure of LHC would be guaranteed 
with $\frac{2}{3}$ probability, and in that case, the chance of total 
failure probability would not change greatly whether we perform 
a card game project or not.
 In that case we would just move some failure probability from the 
``normal'' failure due to the card game or the similar case.

If we place some economical value on the degree of confidence 
we would obtain if our model were indeed true depending on 
whether one failure or another really occurred, 
we could put this benefit into the form 
\begin{eqnarray}
  b_{1)}&&=c\cdot \mathrm{``remarkableness''} \nonumber \\
        &&=c\cdot
        \left\{ \begin{array}{l}
           |ln\,p| \ \mathrm{if\ game\ failure} \\
           |ln\left\langle f \right\rangle\!|=|ln\,f_o|-\frac{\sigma^2}{2}
           \ \mathrm{if\ ``normal''\ failure}. 
        \end{array} \right.  
\end{eqnarray}

In the case of our theory being right, which occurs with probability 
$r$, we estimated that LHC would be stopped with $\frac{2}{3}$
probability \cite{search} so that this benefit will be calculated as an average, 
\begin{eqnarray}
  \left\langle b_{1)}\right\rangle 
     && =c\cdot \mathrm{``remarkableness''} \nonumber \\
     && =c \left\langle \Biggl( \frac{p}{f+p} |ln\,p|+\frac{f}{f+p}
         \left(|ln\, f_o |- \frac{\sigma^2}{2} \right) \Biggr) r\frac{2}{3}
         \right\rangle_{\mathrm{Gauss}},
\end{eqnarray}
where the average $\left\langle \cdots \right\rangle_{\mathrm{Gauss}}$ 
is merely the average over distribution \bref{Gauss} .

 For instance, in the limit of a very small probability $p$ assigned to 
the random number restricting LHC, we would get 
\begin{eqnarray}
\left\langle b_{1)}\right\rangle \approx 
    c\left(|ln\, f_o |- \frac{\sigma^2}{2} \right) r\cdot\frac{2}{3}
    +cp\left\langle \frac{1}{f} \right\rangle \left(|ln\,p|-|ln\,f_o | +
    \frac{\sigma^2}{2}\right) r\> \frac{2}{3}+ \ldots . \label{lowp}
\end{eqnarray}
If, on the other hand, we set $p\gg \left\langle f \right\rangle$, 
we would get
\begin{eqnarray}
\left\langle b_{1)}\right\rangle \approx  \Biggl(|ln\,p|+
\frac{\left\langle f \right\rangle}{p} \left(|ln\,f_o |- \frac{\sigma^2}{2}
-|ln\,p|\right) \Biggr) r\cdot \frac{2}{3}. 
\end{eqnarray}

It is important to notice that, as the previous discussion suggested, 
the correction term in \bref{lowp}
will, for small enough $p$, give increasing benefit with increasing $p$ 
so that it would be beneficial $w. r. t.$ this benefit 
$b_{1)}$ of attaining an increase in the safety of our knowledge 
that $p$ is not completely zero in our model.

\section{Avoiding bad backwardly caused events}

~~~~In our earlier paper, 
we included, in our estimates of whether it would pay to perform our card game 
or random number game experiment, the consideration 
that if we indeed have backward causation for LHC becoming inoperable, 
then these pre-arrangements could have side effects that might be bad 
and, a priori, perhaps also good.
 The backward causation effects might end up being huge in much the same way 
as the famous forward causation effect of the butterfly in the 
``butterfly effect'', but in the same way as it is difficult to predict 
whether the effects are good or bad when the butterfly beats its wings 
in a particular way, 
it is hard to know if the pre-arrangements set up to prevent LHC  
from working are good or bad.
 If we think of such possibilities as the closure of CERN or an earthquake 
in Geneva, 
we may judge it to be bad, but if we think of even earlier or further distant 
pre-arrangements, it becomes increasingly difficult to estimate 
either good or bad.
For instance, it is a possibility that a major factor behind the SSC
being terminated by Congress was 
the collapse of the Soviet Union \cite{13}. 
 This were a huge backward causation effect but it is hard to evaluate the probability 
as to whether it is good or bad. 
 Thus, it would have been hard to evaluate, in advance, whether 
our card game would have been profitable had our theory been known then.

In the previous articles \cite{search}, 
we called the price of the damage arising in excess 
when a ``normal'' failure of LHC is provoked, $d$ . 

We should imagine that the very huge backward causation effects occurring 
very remotely  
from the LHC are probably averaged out to zero, 
similar to the far future effects of the butterfly wing. 
 Hence the important contributions to the damage cost 
 $d$ are rather close in time (and space) to the LHC itself. 
 We very roughly estimated, in our previous study, 
$d \approx 10 \cdot $ ``cost of LHC'' $\simeq 10 \cdot 3.3 \cdot 10^9$ CHF
$=3.3 \cdot 10^{10}$ CHF. 

 In the case of the card game failure, there may also be huge effects, 
but now the evaluation of the damage being good or bad would be totally 
opaque.
 Only the effects of performing the actual experiment may have any predictable 
average effect. Therefore, in the case of such an artificial failure, 
the damage would be limited \blue{to} statistically washing out damage 
(i.e., they are equally likely to be good or bad) and the obvious loss 
because of the  restriction on the  
$d_\mathrm{\,rest.loss}$ card drawn. 

We should arrange the latter damage to almost certainly be the minimal 
one by assigning mild restrictions to be much more likely outcomes than 
heavy restrictions. 

The damage done, or by switching the sign, the (negative) benefit, is 
\begin{eqnarray}
-b_{2)}=d\cdot\frac{2}{3}r\cdot \frac{f}{f+p}+d_\mathrm{\,rest.loss} \cdot
\Biggl( p\left(1-\frac{2}{3}r\right)+\frac{2}{3}r\cdot \frac{p}{f+p}\Biggr) , 
\end{eqnarray}
where we used the notation $d_\mathrm{\,rest.loss}$ for the cost of the restrictions. 

\section{Conclusion and outlook}

We have discussed in this article two major topics in connection with
our previously proposed model with the action assumed to be complex.

The first of these subjects could have been considered starting from the 
troubles of EPS problem 
the Bell's theorem which states that quantum mechanics makes predictions
in the case of entangled particles being measured on that are in 
disagreement with seemingly very reasonable assumptions.  There is however 
as noticed by Bell himself a way out for quantum mechanics if one
makes use of that for given initial conditions the measurements which the
experimentalists at the two discussed significantly separated positions A 
and B in the Bell or EPS experiment perform is already in principle 
determined by determinism of at least say classical approximation physics.
This deterministically determined choice of experiment being performed
namely makes the need for discussing simultaneously several possible 
choices ( by ``free will'' so to speak) irrelevant. In our complex action 
model this point may be more stressed since the initial conditions are 
even in principle calculable.

However, we believe that it is NOT this true superdeterminism which
makes our model with the complex action more able to cope with the
Bell's theorem problem, but rather the fact that our model predicts that
\emph{the measurement results depend on the happening in the future}!
It is this backward causation property of our model which makes
the assumption of no signal going faster than the speed of light
being a prerequisite for Bell's theorem not trustable in our model.
The point is that if the future can influence the past by making an 
adjustment of the initial conditions or by as is here relevant influence
the outcome of a measurement, then a genuine signal coming along
with less than speed of light from A to B is not needed. Instead we can 
have an effect from the future which is influenced by a signal from
A. But if one can wait to get the signal to somewhere in the future of 
course there is no need for the signal reaching along faster than light. 
It has time enough to reach the future, just influence can go backward in 
time there is no hurry to get the signal along.
 Actually we found that our model essentially reproduces in a second quantized 
version in principle calculable classical fields which can be identified 
with the wavefunctions including echoes from future in the transactional 
interpretation.
To our model is with respect to the essential picture identical to the transactional 
interpretation model, although we do not have exatly the Feynman-Wheeler 
time reflection invariant emission exactly. Rather our influence from future 
is determined by parameters in the imaginary part of the action.

Thus we claimed that actually the Bell's trouble calls for the influence
from future effect, and thus one should really attempt to look for such
backward in time influences whenever some new region of physics is being 
explored. Using our special model of complex action the obvious place to 
look for such effects namely the at a given time highest energy 
accelerator gets especially suggested. So we should look for such effects 
by means of LHC.

We have argued that it would be a good idea to perform 
our earlier proposed experiment of generating some random numbers 
-- by card drawing or by a quantum random number generator, 
or even both ways -- 
and letting them then be decisive in applying restrictions on the beam energy and/or 
the luminosity and/or the like. 

The main point was that our theory, referred to as 
``model with an imaginary part of the action'', 
is indeed seen to be right if LHC is 
stopped by our proposed game rather than if it just failed for some technical 
or political reason. 
The reason for the suggestion of our model being right if the LHC were 
stopped by a random number (card) game decision than by just a 
``normal" technical or political failure is that it is very hard to estimate 
in advance how likely it is for a ``normal'' failure of LHC to occur. 

The greatest encouragement for performing the experiment without much hesitation 
is the remark that  
whatever happens with our proposed experiment, it will, in practice, seem to 
be a success or at least to be of no harm. 
The point is that in the case of any restriction being imposed 
by the random numbers, we have, because of the very fact of 
these random numbers being generated at all, obtained the shocking 
great discovery that there is ``backward causation.'' 
Such a discovery of the future influencing the present and past would 
be monumental. 
Consequently, we would be very happy and it would be a fantastic success 
for the LHC to have caused such a discovery!
\renewcommand{\thesection}{}
\section{Acknowledgements}

~~~~The authors acknowledge the Niels Bohr Institute and 
Okayama Institute for Quantum Physics for the hospitality 
extended to them. This work was supported by the Grants -- 
in-Aids for Scientific Research, No. 19540324,
and No. 21540290, from the 
Ministry of Education, Culture, Sports, Science and Technology, Japan. 




\begin{thebibliography}{99}
\bibitem{search}
  H.~B.~Nielsen and M.~Ninomiya,
    ``Test of Effect from Future in Large Hadron Collider; 
    A Proposal", arXive:0802.2991 [physics, gen-ph] 
    IJMPA vol. 24, No. 21-22, PP. 3945-3968 (2009); ibid.
    ``Search for Future Influence from LHC".
IJMPA  vol. 23, Issue 6, PP. 919-932 (March 10, 2008);
arXiv:0707.1919 [hep-ph].
 
\bibitem{3}
 H.~B.~Nielsen and M.~Ninomiya,
``Future Dependent Initial Conditions from Imaginary Part in Lagrangian",
Proceedings of the 9th Workshop 
``What Comes Beyond the Standard Models?",
Bled, September 16-26, 2006, 
DMFA Zaloznistvo, Ljubljana;
 hep-ph/0612032.

\bibitem{4} H.~B.~Nielsen and M.~Ninomiya, ``Law Behind Second Law of 
Thermodynamics-Unification with Cosmology'', JHEP, 03,057-072 (2006);
 hep-th/0602020.

\bibitem{5}
 H.~B.~Nielsen and M.~Ninomiya,
 ``Unification of Cosmology and Second Law of Thermodynamics:
 Proposal for Solving Cosmological Constant Problem, and Inflation",
Prog. Theor. Phys., {\bf Vol. 116, No. 5} (2006);
  hep-th/0509205, YITP-05-43, OIQP-05-09.

\bibitem{r1}
 J. Faye,
``The Reality of the Future",
Odense University Press.

\bibitem{6}
 J.~S.~Bell: He never published a written work on 
 ``superdeterminism", but he talked about it on BBC Radio (1985). 
 Some quotes from his talk on BBC are found in the web-page \\
http://groups.google.com/group/sci.bio.technology/browse\_thread/thread/\\f76bf7af7b11097f
written by ``Archimedes Plutonium" (1985).

\bibitem{Cramer} 
J.~G.~Cramer, 
``The transactional interpretation of Quantum Mechanics", 
Rev. of Mod. Phys. vol. 58, PP. 647-688 (1986).  
 
\bibitem{Wheeler_Feynman_interaction} 
J.~A.~Wheeler and R. P. Feynman,  
``Interaction with the Absorber as the Mechanism of Radiation", 
Rev. of Mod. Phys. vol. 17, PP. 157-161 (1945);
J.~A.~Wheeler and R.~P.~Feynman, 
``Classical Electrodynamics in Terms of Direct Interparticle Action", 
Rev. of Mod. Phys. vol. 21, PP. 425-433 (1949)  

\bibitem{7} 
D.~L.~Bennett, C.~D.~Froggatt and H.~B.~Nielsen, 
``Nonlocality as an Explanation for Fine Tuning in Nature", 
in Proceedings of 28th International Symposium on Particle Theory, \,
30 Aug.\,3 Sep. 1994,\, Wendisch-Rietz, Germany, PP. 394-412. 

\bibitem{8}
D.~L.~Bennett, C.~D.~Froggatt and H.~B.~Nielsen, 
``Nonlocality as an Explanation for Fine Tuning 
  and Field Replication in Nature", 
arXive hep-ph/9504 294, in the Adriatic Meeting on Particle Physics: 
Perspectives in Particle Physics `94,13-20 Sep. 1994 Island of Brijuni, 
Croatia, PP. 255-279. 

\bibitem{9}
D.~L.~Bennett, ``Multiple Point Criticality, Nonlocality, and Finetuning in Fundamental Physics: Predictions for Gauge Coupling Constants gives $\alpha^{-1}=136._8\pm 9$", 
Ph. D. Thesis, arXive:hep-ph/9607341.

\bibitem{10}
H.~B.~Nielsen and C.~Froggatt, ``Influence from the Future", 
arXive:hep-ph/9607375, in Proceedings of 5th Hellenic School 
and Workshops on Elementary Particle Physics, Corfu, Greece, 
3-24, Sep. 1995. D.~L.~Bennett, ``who is Afraid of the Past'',
talk given by D.~L.~Bennett at the meeting of the Cross-displiary 
Initiative at Niels Bohr Institute on September 8, 1995. QLRC-95-2. 

\bibitem{14}
H.~B.~Nielsen, ``Imaginary part of action, future functioning as hidden variables", 
in Proceedings  o the conference QTRF-5, June 14-18, 2009. 
Vaxjo Sweden, arXive:0911.4005V1 [Quantum Ph]

\bibitem{11}
I.~Stewart, ``An Iterated Search for Influence from the future on the Large Hadron Collider", arXive:0712.0715v2 [hep-ph], 15 Dec. 2007. 

\bibitem{12}
J. Mervis and C. Seife,
10,1126/science, 302.5642.38,
``New Focus: 10 Years after the SSC.
Lots of Reasons, but Few Lessons".

\bibitem{13}
M.~Riordan, ``The Demise of the Superconducting Super Collider'', Phys. Perspect vol. 2, 
PP. 411-425 (2000)

\end{thebibliography}
\end{document}